**Title**

**Non-Hermitian Singularities in Scattering Spectra of Mie Resonators**

**Authors**

Fan Zhang,[1,2]† Nikolay S. Solodovchenko,[2]† Hangkai Fan,[3] Mikhail F. Limonov,[2,4] Mingzhao Song,[1]* Yuri S. Kivshar,[1,5]* Andrey A. Bogdanov[1,2,4]*

**Affiliations**

[1] Qingdao Innovation and Development Center, Harbin Engineering University, Qingdao, 266000, Shandong, China.

[2] School of Physics and Engineering, ITMO University, St. Petersburg 191002, Russia.

[3] College of Information and Communication Engineering, Harbin Engineering University, Harbin, 150001, Heilongjiang, China.

[4] Ioffe Institute, St. Petersburg 194021, Russia.

[5] Nonlinear Physics Center, Australian National University, Canberra ACT 2601, Australia.

* Corresponding author.

Email: kevinsmz@foxmail.com (M.S.); yuri.kivshar@anu.edu.au (Y.S.K); bogdan.taurus@gmail.com (A.A.B.)

† These authors contributed equally to this work.

**Abstract**

Non-Hermitian systems are known to possess unique singularities in the scattering spectra such as exceptional points, bound states in the continuum, Diabolic points, and anapole states, which are usually considered to be independent. Here, we demonstrate the fundamental relationships between non-Hermitian singularities and observe them experimentally in the scattering spectra. We reveal that exceptional points appear in the anapole regime, and diabolic points are associated with superscattering. We confirm our findings with microwave experiments by measuring the scattering spectra of subwavelength Mie-resonant ceramic rings. Our study underpins the generic behavior of non-Hermitian singularities in the scattering spectra of subwavelength resonators, uncovering their novel applications in non-Hermitian nonlinear optics and topological photonics.

**Teaser**

Exploring the non-Hermitian singularities and their links in the scattering spectra of Mie resonators reveals new paths for field manipulation and control.

**MAIN TEXT**

# Introduction

    The study of Mie resonances in subwavelength dielectric particles has significantly advanced our understanding of the scattering and localization of electromagnetic waves (*1,2*). In particular, the subwavelength resonators can support multipole electric and magnetic Mie resonances, and their control and engineering allow manipulating light-matter interaction at a subwavelength scale. Each Mie resonator is an open physical system that, due to radiative losses, is characterized by complex eigenvalues, and thus it becomes an object of non-Hermitian physics (*3*). Complex eigenvalues can be manipulated through the meticulous tailoring of the structural parameters, thereby enabling a variety of exotic phenomena driven by coupling and interference between multipolar resonances, including quasi-bound states in the continuum (Q-BICs) which theoretically possess ultrahigh quality factors and the near-field localization (*4-6*), the unidirectional scattering Kerker effect (*7*), anapole states characterized by the scattering dark state (*8*), super-scattering (*9*) and super-multipole (*10*) which can break the limit of single-channel scattering.

    Non-Hermitian physics is known to offer many distinctive phenomena that are not accessible in Hermitian systems (*11*). In particular, exceptional points (EPs) are unique singularities in non-Hermitian systems where the eigenvalues and eigenvectors coalesce (*12-15*). These singularities give rise to various novel physical phenomena, such as chiral mode switching (*16-18*), unidirectionality (*19-21*), nonreciprocal wave propagation (*22*), coherent perfect absorption (*23-25*), and enhanced sensitivity (*26,27*), making them a focal point in numerous disciplines, including acoustics (*28*), mechanics (*29*), electronics (*30*), semiconductor physics (*31,32*), photonics (*33-35*) and plasmonics (*36-40*). EPs are closely associated with the concept of parity-time (PT) symmetry, providing the critical link between symmetric and broken PT phases. Although traditionally associated with gain-loss balance, recent studies have shown that EPs can also be realized in passive systems exemplified by their application in highly sensitive protein detection in plasmonic systems (*41*). Especially, EPs originating from Mie resonances have been studied only theoretically (*42,43*), and these studies are limited by the analysis of eigenvalues and evolution of the scattering spectrum. The unique features of non-Hermitian wave scattering from Mie resonators were never demonstrated in the experiment.

    From the perspective of multipole channel coupling, it is essential to elucidate different phenomena associated with Mie resonances and non-Hermitian singularities. Radiative coupling can be decomposed into multipolar radiation, and by examining the non-Hermitian systems from a multipole perspective, we can gain an understanding of the effect of complex

eigenmodes. In this paper, we numerically and experimentally study non-Hermitian Mie singularities. By an example of a dielectric ring-shaped resonator, we explore the transition between EPs and DPs by manipulating structural parameters to control the system detuning and mode coupling and to observe and utilize EPs. The primary objective of this study is to elucidate the dynamics of non-Hermitian singularities in dielectric ring resonators and to demonstrate their potential applications in controlling and enhancing light-matter interactions. By understanding the underlying mechanisms and characteristics of EPs and DPs, we aim to pave the way toward the development of highly tunable sensors and other optical devices that leverage these unique non-Hermitian properties. We believe our results not only contribute to the fundamental understanding of the insight physics of non-Hermitian singularities in Mie scattering but also open up new possibilities for potential applications in various optical technologies. The ability to manipulate and utilize the scattering behavior of non-Hermitian singularities in dielectric resonators promises to revolutionize the design and functionality of next-generation photonic devices.

**Results**

As an example, we consider the structure illustrated in **Fig. 1A** in the form of a dielectric ring-shaped subwavelength Mie resonator with permittivity ($\varepsilon_r$) of 80 placed in the air ($\varepsilon_0 = 1$), which brings a realization of a non-Hermitian system. For theoretical investigation of the singularities in such a non-Hermitian system, we consider an open system model based on the coupled modes with different symmetry, depicted in **Fig. 1B**. The model assumes the existence of three distinct modes, the first two of which exhibit the same symmetry and thus near-field and far-field coupling. The last mode, however, exhibits different symmetry, which precludes any interaction due to symmetry prohibitions, ensuring the orthogonality between these modes. The theoretical framework employed to describe this system is the Temporal Coupled Mode Theory (TCMT), and the dynamics of the open system can be described using a Hamiltonian matrix:

$$\widehat{H} = \begin{pmatrix} \omega_1 & \kappa & 0 \\ \kappa & \omega_2 & 0 \\ 0 & 0 & \omega_3 \end{pmatrix} - \begin{pmatrix} \gamma_1 & e^{i\psi}\sqrt{\gamma_1\gamma_2} & 0 \\ e^{i\psi}\sqrt{\gamma_1\gamma_2} & \gamma_2 & 0 \\ 0 & 0 & \gamma_3 \end{pmatrix} \quad (1)$$

where the first term is the Hermitian part describing a lossless system, and the second term is the anti-Hermitian part describing radiation of two resonances with the same symmetry. $\omega_j$ and $\gamma_j$ (j = 1,2,3) represent the resonant frequencies and leakage rates of the resonances. Parameter $\kappa$ denotes the near-field coupling strength between two symmetric modes, $e^{i\psi}\sqrt{\gamma_1\gamma_2}$ is the interference of the radiative waves through far-field coupling, $\sqrt{\gamma_1\gamma_2}$ is the radiative coupling term, while $\psi$ is the phase difference between two symmetric modes. The complex eigenvalues of the Hamiltonian (1) are:

$$\begin{cases} \widetilde{\Omega}_\pm = \omega_{\text{ave}} - i\gamma_{\text{ave}} \pm \sqrt{(\Delta\omega - i\Delta\gamma)^2 + \left(\kappa - e^{i\psi}\sqrt{\gamma_1\gamma_2}\right)^2} \\ \widetilde{\Omega}_3 = \omega_3 - i\gamma_3 \end{cases} \quad (2)$$

where $\widetilde{\Omega}_\pm$ represents the complex eigenvalues induced by the modes with same symmetry where the coupling is considered, $\widetilde{\Omega}_3$ represents the complex eigenvalue of the mode with opposite symmetry without coupling. $\omega_{\text{ave}} = (\omega_1 + \omega_2)/2$ and $\gamma_{\text{ave}} = (\gamma_1 + \gamma_2)/2$ are the averages of $\omega_j$ and $\gamma_j$ (j = 1,2), while $\Delta\omega = (\omega_1 - \omega_2)/2$ and $\Delta\gamma = (\gamma_1 - \gamma_2)/2$ are the differences of $\omega_j$ and $\gamma_j$ (j = 1,2), respectively.

Learned from equation (2), the complex eigenvalues can be freely tailored by manipulating the detuning and coupling regardless of the structure. As an example of the proposed single dielectric ring resonator, one of the simplest ways to control the detuning and coupling is varying the structural parameters [i.e. aspect ratio $(R_{\text{out}} - R_{\text{in}})/h = \Delta R/h$, and radius ratio $R_{\text{in}}/R_{\text{out}}$]. The external radius $R_{\text{out}}$ is fixed to guarantee that only two degrees of freedom in the proposed system. **Fig. 1C** shows the critical coupling strength ($R_{\text{in}}/R_{\text{out}} = 0.4983$) by changing $\Delta R/h$, EP appears at the critical state characterized by the coalescing of both real and imaginary parts of the eigenvalues indicated by the green circles [**Fig. 1C (IV)**]. DP appears at the crossing point where only the real eigenvalues merge indicated by the yellow circle [**Fig. 1C (VI)**]. At $\Delta R/h$ = 0.4524, The electric field distributions of the two eigenmodes are the same [**Fig. 1B (IV)**], since the coalescence of the eigenvector is also at EP. While at $\Delta R/h = 0.4865$, The electric field distributions of the two eigenmodes are different, since the associated eigenvectors are orthogonal [**Fig. 1B (VI)**]. In **Fig. 1D**, we explain the scattering behavior at EP and DP associated with multipole Mie resonances. EP is related to the dark scattering state which meets the anapole condition, while DP is related to the super scattering state which meets the Kerker condition.

To demonstrate the existence of the EP in above system and characterize the scattering behavior, we consider to plot the scattering cross-section maps of a lossless dielectric ring-shaped resonator with varying $\Delta R/h$ in simulation, shown in **Fig. 2C**. The dashed white lines indicate three types of the simulated eigenvalues which are the same with **Fig. 1C**. Two crossing points (i.e. EP and DP) can be found but locate at dip and peak of the scattering cross section (SCS), respectively. **Fig. 2D** shows the induvial SCS curves at EP [along **Fig. 2C(I)**] and DP [along **Fig. 2C(II)**]. At $\Delta R/h = 0.4524$, EP is located on the dip in the SCS between two resonance peaks, a phenomenon that is similar to coupled-resonator-induced transparency and electromagnetically induced transparency. At EP, two complex eigenvalues (blue and red) coalesce in the eigenvalue space. At $\Delta R/h$ = 0.4865, DP is situated at the overlapping of resonance peaks, which is analogous to the superscattering state. At DP, only the real part of two complex eigenvalues (gray and red) merge.

For the experimental verification, we fabricate a set of dielectric ring-shaped samples (the external radius $R_{out}$ = 1.6 cm, $R_{in}/R_{out}$ = 0.498) with a varying height, the selected compound material is BaO-Ln$_2$O$_3$-TiO$_3$ with a permittivity $\varepsilon$ = 80.5 and loss tangent $\delta$ = $1.8 \times 10^{-4}$. Details about the sample fabrication and far-field experimental setup can be found in Materials and Methods. The measured SCS for the normal incident wave is plotted in **Fig. 2E** and shows good agreement with the numerical simulation results in the upper panel, with differences due to the scattering from the inhomogeneity, fabrication defect and the uncertainty of the measurement environment. Experimental fitting results (white solid lines of **Fig. 2E**) show a tiny displacement of the two crossing points of Re($\widetilde{\Omega}_\pm$), at approximately $\Delta R/h$ = 0.452 and $\Delta R/h$ = 0.488, due to the precision limits of sample preparation. In **Fig. 2F**, we investigate the evolution of SCS results from the experiment (purple solid lines) with the corresponding complex eigenvalues at different $\Delta R/h$. At $\Delta R/h$ = 0.452, two complex eigenvalues (blue and red) are close, indicating the presence near the EP. At $\Delta R/h$ = 0.488, two complex eigenvalues (gray and red) degenerate only in the real part, representing close to DP.

To more accurately obtain more information from experimental curves, we consider a method based on quasi-normal modes (QNMs) (*44*), which allows us to obtain the scattering spectrum from each eigenmode separately. The response from each QNM is fitted by using the Fano formula and all the necessary resonance parameters are obtained: Fano parameter $q$, intensity $I$ and eigenfrequency $\omega_j - \gamma_j$ (j = 1,2,3). The data obtained are employed as an initial approximation of parameters for the purpose of fitting experimental spectra, with the objective of obtaining frequencies and radiation losses. **Fig. 2G** and **Fig. 2H** compare the simulation and experimental normalized complex eigenvalues with different $\Delta R/h$. (with a detailed comparison between fitted and measured SCS curves in Supplementary Material). The excellent agreement shows the validity of this QNMs method. The difference between the numerically calculated complex eigenvalues and the experimental results mainly stems from the scattering from the sample surface roughness and inhomogeneity in fabrication.

Learning from the behavior of scattering spectra, we understand that EP appears in the dip between two resonance peaks [**Fig. 2D(I)**/ **Fig. 2F(I)**] while DP appears in the overlap of two resonance peaks [**Fig. 2D(II)/ Fig. 2F(II)**]. To gain a deeper insight into the scattering behavior of these singularities, we have employed multipole decomposition. **Fig. 3(A-F)** illustrate the first three orders of multipoles for the SCS of the simulation in **Fig. 2C**. The ring-shaped dielectric resonator exhibits inversion symmetry, which results in the eigenmodes of the resonator comprising either only odd (blue dashed frame in the right panels of **Fig. 3**) or even (red dashed frame in the right panels of **Fig. 3**) multipoles (*45-47*). **Fig. 3A-3C** illustrate the primary multipoles of odd modes. Two odd modes (white circles and squares), which are predominantly attributed to the electric dipole, are observed to degenerate at the position of EP. **Fig. 3G** illustrates the proportion of multipoles at EP of three modes. The destructive interference between two electric dipoles (ED) contributed odd modes (mode 1 and mode 2) in the far-field results in the scattering dark state (i.e. anapole state) in SCS between the position of two eigenmodes, as illustrated in **Fig. 3A** by the black solid lines. It is also worth mentioning that Q-BIC appears during the multipolar conversion from dipole to quadrupole (mode 2) shown in **Fig. 3I**. A comparable phenomenon is also observed in two magnetic dipoles (MD), as shown in the Supplementary Material. **Fig. 3D-3F** show the main multipoles of even modes (white diamonds). The DP is observed at the crossing position where an odd mode (mode 1) merges with an even mode (mode 3). This phenomenon can be described as the superposition of ED and MD, as shown in **Fig. 3H**. Since ED and MD are not subject to interference in the far-field, the scattering behavior in the SCS can be considered as the superposition of resonance peaks. Further details on the quantification of multipoles at different singularities can be found in the Supplementary Material.

The location of far-field scattering suppression tends to result in a pronounced near-field enhancement effect (especially anapole state), and therefore we consider conducting near-field detection experiments on the proposed ring resonator. Details about the near-field experimental setup and sketch can be found in **Fig. 4(A-B)** and Materials and Methods. Firstly, a numerical study is conducted on the normalized SCS (red line in **Fig. 4C**) and the y-component of the near-electric-field enhancement in the center position inside the ring-shaped resonator (black line in **Fig. 4C**). It is observed that the position of the dip in the normalized SCS exhibits a slight difference to the maximum of the near-electric-field enhancement. Furthermore, the y-component of the electric field exhibited a comparable profile, albeit with varying degrees. Consequently, the optimal position can be identified as that which achieves the greatest near-electric-field enhancement. Additionally, we also investigate the evolution of field enhancement by varying the observation plane height to choose the best height for the experiment. As illustrated in **Fig. 4D**, a height of approximately 3 mm proved to be an optimal range for achieving field enhancement experimentally, with minimal attenuation on near-electric-field enhancement. In **Fig. 4E**, the $R_{in}/R_{out}$ of the sample is fixed at 0.4985, and the $\Delta R/h$ is uniformly changed from 0.42 to 0.47. This allows for straightforward control of the near-electric-field strength enhancement, which provides an effective method for enhancing and controlling light-matter interference. The experimental results are presented in **Fig. 4F**, where the colorful dashed lines demonstrate a satisfactory correlation with the simulated results, as indicated by the corresponding colorful solid lines shown in **Fig. 4E**. At 3mm above the surface of the ring resonator, we investigate the y-component of electric field and z-component of magnetic field in both the simulation (**Fig. 4G**) and the experiment (**Fig. 4H**). It can be observed that there is a pronounced electric-field enhancement in the central surface of the resonator.

We further investigate the sensitivity with varying structural parameters, and discuss the system robustness of the sensitivity in the vicinity of EP. **Fig. 5A** illustrates the evolution of Re($x$) with varying $\Delta R/h$ and $R_{in}/R_{out}$, where the golden star indicates the position of EP. We consider two directions (D1 and D2) to demonstrate the robustness of the sensitivity at EP. D1 (yellow arrow) means fixed $R_{in}/R_{out}$ and varying $\Delta R/h$, D2 (blue arrow) means optimizing $R_{in}/R_{out}$ of every corresponding $\Delta R/h$ along the PT-symmetry lines (in accordance with Supplementary Table). **Fig. 5B** demonstrates the simulated response to the perturbation, represented as the splitting degree of the eigenfrequency, of EP (blue and yellow) and DP (red) as histograms, while the scatter plots represent the experimental results. The red bar indicates the degree of response to changes in $\Delta R/h$ at DP. The red diamond in the inset of **Fig. 5C** indicates that the slope is 1, implying that the response to external perturbation is linear at DP sensing. When along the D1, the yellow bar displays the degree of resonance splitting at EP. The slope of yellow squares in the inset of **Fig. 5C** is 0.75, indicating that the limit of EP sensing has not been reached. When along the D2, the blue bar graph illustrates the degree of resonance splitting degree at EP when the PT-symmetry condition is met. The blue circles in the inset of **Fig. 5C**, which exhibit a slope of 0.5, indicate a square-root relationship. This implies that the maximum theoretically achievable nonlinear response at EP sensing has been reached. As illustrated in **Fig. 5C**, the maximum degree of response to perturbation at EP is more than five times higher than that at DP, according to theoretical calculations.

**Discussion**

We have uncovered the power of non-Hermitian singularities in the Mie scattering by studying the behavior of scattering around EPs and DPs through meticulous system detuning. By carefully actuating the height of a subwavelength dielectric ring, we have detected unique eigenvalue behaviors and field distributions and observed a distinct state transition between scattering dark states at EP and superscattering states at DP. Our study of non-Hermitian singularities highlights novel opportunities for developing advanced optical devices with enhanced sensitivity and control over light-matter interactions. Our results not only pave the way towards the development of compact and functional metadevices (such as sensors, absorbers, lasers, and energy harvesting/transmission units) with high sensitivity and tunable quality factors but also unveil unexplored avenues within the dynamic landscape of multipolar Mie resonances in non-Hermitian photonics.

.

**Materials and Methods**

**Numerical simulations**

The full wave simulations to calculate both the eigenvalues and the scattering cross section are performed in the Wave Optics module of COMSOL Multiphysics, a commercial finite element simulation software. The dielectric resonator is surrounded by an air layer (more than one wavelength thick) and a 10-layer perfectly matched layer (PML). A physically controlled mesh with extremely fine element size is used. Since the dielectric ring is inversion symmetric, we reduce the 3D model to a 2D axisymmetric modal, following the method introduced in (*45*). The electromagnetic fields can be expanded into a Fourier series of waves corresponding to different azimuthal indices. For the calculation of eigenvalues, the azimuthal number is set to 1. To calculate the scattering cross section, the results are summed with the azimuthal number from -3 to 3, which can almost be considered as the total scattering cross-section.

**Description of the experiment details**

The image of the sample is shown in the inset of **Fig. 2A**, placed on a foam plate holder which can be considered transparent at microwave frequency band. We perform the far-field experiment in an anechoic chamber over the microwave frequency band 2-3 GHz. The photo and sketch of the experimental setup to measure the SCS of the ring-shaped resonator are shown in **Fig. 2A** and **Fig. 2B**. We choose a broadband horn antenna generating the normal incident plane wave as a transmitter to excite the sample and another antenna with the same configuration as a receiver, forming a two-port network. The distance $d \approx 2m$ which is 10 times than the working wavelength, making the far-field condition. The sketch depicted in **Fig. 2B** illustrates the same experimental configuration of the experimental setup as shown in **Fig. 2A**. A vector network analyzer (VNA) is utilized to generate microwave and collect transmission spectra. Two horn antennas serve as normal incident microwave transmitter and receiver respectively with linear polarization. The laser is utilized to collimate the horn antennas and measured samples. Before measuring transmission spectra, the VNA is calibrated without a sample to subtract the background.

The scattering cross section of the single dielectric resonator is measured in an anechoic chamber. The total scattering cross section was calculated from the measured complex transmission coefficient $S_{21}$(49,50):

$$\sigma_{ext}(f) = \frac{c}{\sqrt{\pi}f} \cdot Im\left(\frac{S_{21}}{S_{21,bg}}\right) \tag{3}$$

where $\sigma_{ext}$ is the extinction cross-section, $c$ is the light speed in the vacuum, $S_{21}$ is the complex transmission coefficients of the resonator with background, $S_{21,bg}$ is the complex transmission coefficients of the background.

**Fig. 4A** and **Fig. 4B** depict the near-field experimental setup and sketch utilized to quantify the near-electric-field intensity. The ring-shaped samples are excited by a normal incident plane wave, and the probe for measuring the near-electric field is positioned above the top surface of the sample. A VNA is utilized to generate microwave and collect the y-component of the near-electric-field distribution. One horn antenna is served as a normal incident microwave transmitter with linear polarization. The probe is $s \approx 3\ mm$, which is the depth of the probe, above the surface of the resonator.

**Acknowledgments**

**Funding:**

National Natural Science Foundation of China (Project No. 62101154)

Natural Science Foundation of Heilongjiang Province of China (Project No. LH2021F013)

Russian Science Foundation (Project 23-72-10059).

Australian Research Council (Grant DP210101292)

**Author contributions:**

    Conceptualization: AAB, FZ, HF

    Methodology: FZ, NSS, AAB

    Investigation: NSS, FZ

    Visualization: FZ, HF

    Supervision: MS, YSK, AAB

    Writing—original draft: FZ

    Writing—review & editing: NSS, HF, MFL, MS, YSK, AAB

**Competing interests:**

Authors declare that they have no competing interests.

**Data and materials availability:**

All data are available in the main text or the supplementary materials.

**Figures and Tables:**

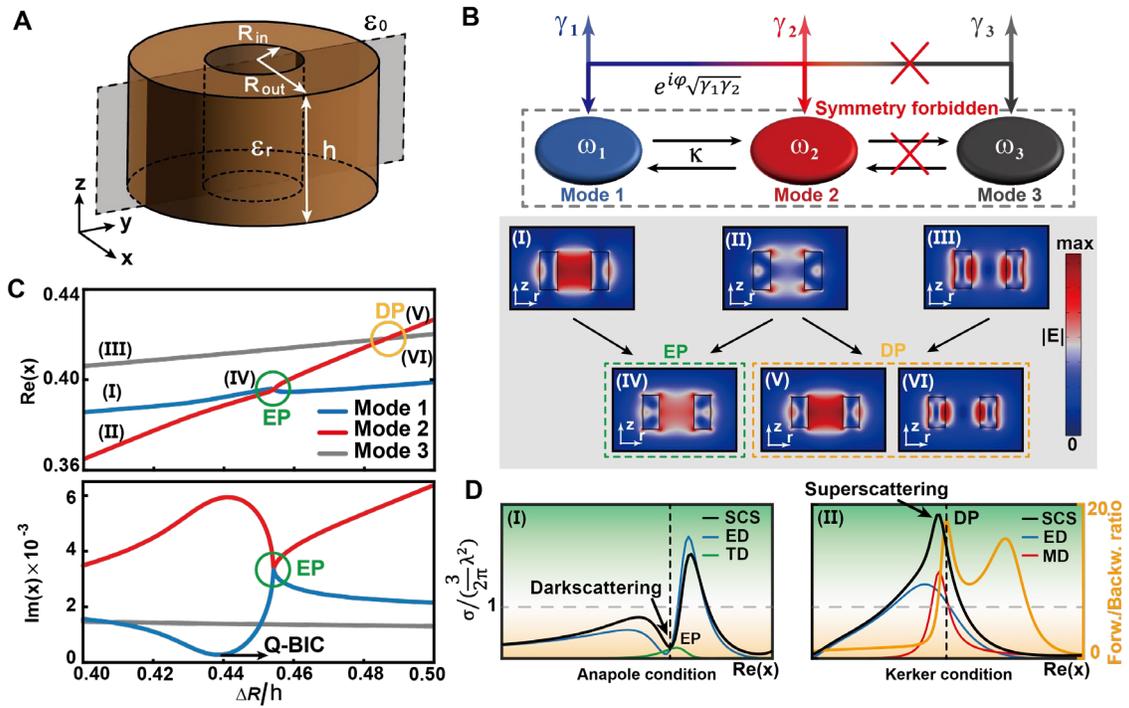

**Fig. 1. Non-Hermitian singularities in a single ring-shaped dielectric resonator. (A)** Schematic illustration of the proposed ring-shaped dielectric resonator. **(B)** Simulated complex size parameter $x = \omega \Delta R/c = 2\pi f \Delta R/c$ of three eigenmodes as the functions of varying aspect ratio $\Delta R/h$. EP can be predicted at the position where both real and imaginary parts of $x$ coalesce ($\Delta R/h$ = 0.4524), while DP appears when only real part of $x$ merge ($\Delta R/h$ = 0.4865). **(C)** General concept of the TCMT describing the proposed non-Hermitian system, which is consist of the closed system circled by the dashed black line and the radiative losses. The electric-field distributions along the cross-section in the gray plane of **(A)** are corresponding to the eigenmodes of **(B)**, where eigenstates coalesce at EP. **(D)** The scattering behavior of anapole state at EP and super scattering state at DP.

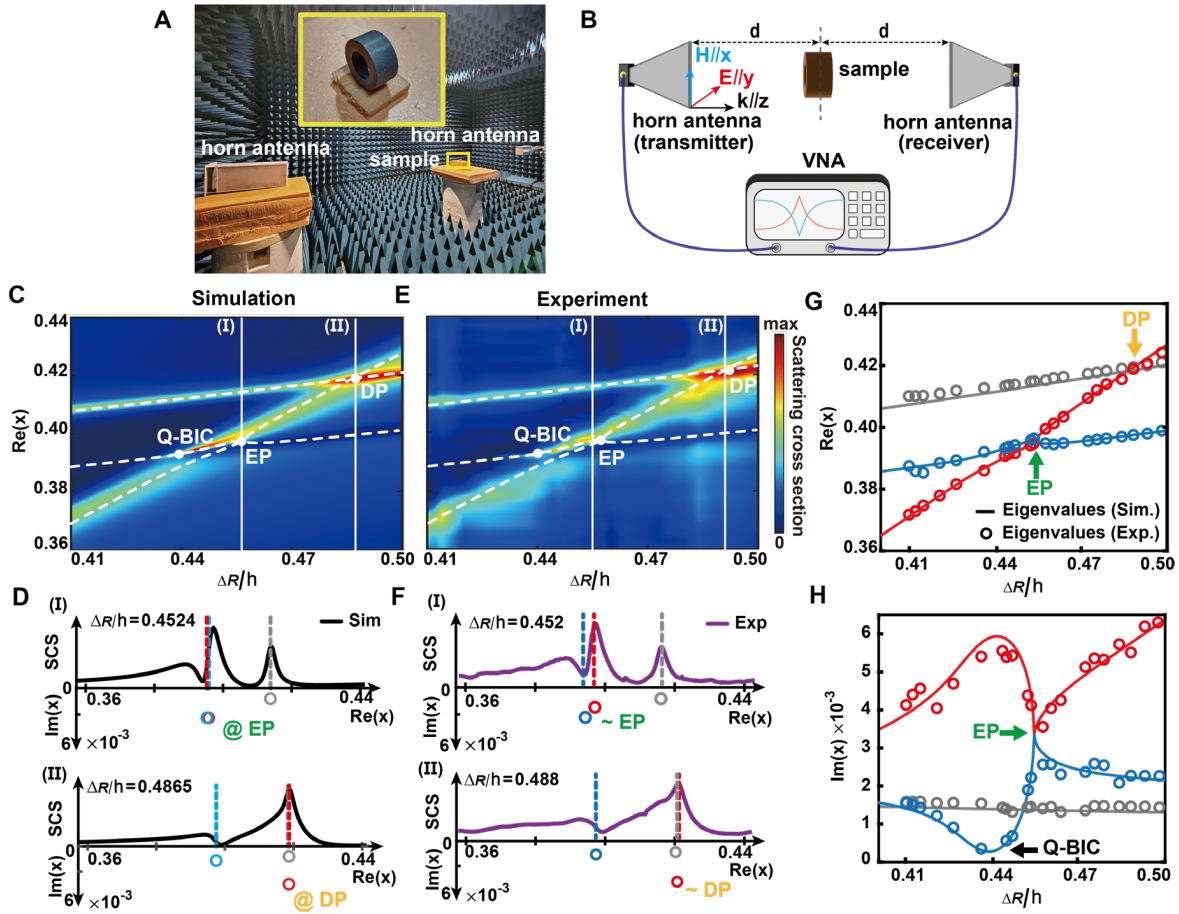

**Fig. 2. Far-field numerical and experimental results. (A)** Far-field experimental setup and **(B)** schematic illustration. **(C)** Simulated map of the scattering cross section under the normal incident wave with the dependency of $\Delta R/h$ and $x$. The white dashed lines are visual auxiliary lines of the eigenvalues. The white dots indicate the position of EP and DP. **(D)** The panels of the simulated scattering cross section at EP and DP. **(E)** Experimental map of the scattering cross section under the normal incident wave with the dependency of $\Delta R/h$ and $x$. **(F)** The panels of the experimental scattering cross section near EP and DP. **(G-H)** The real **(G)** and imaginary **(H)** parts of the simulated and experimental eigenvalues. The colorful solid lines represent the simulated eigenvalues while colorful dots represent extracted eigenvalues from experimental results.

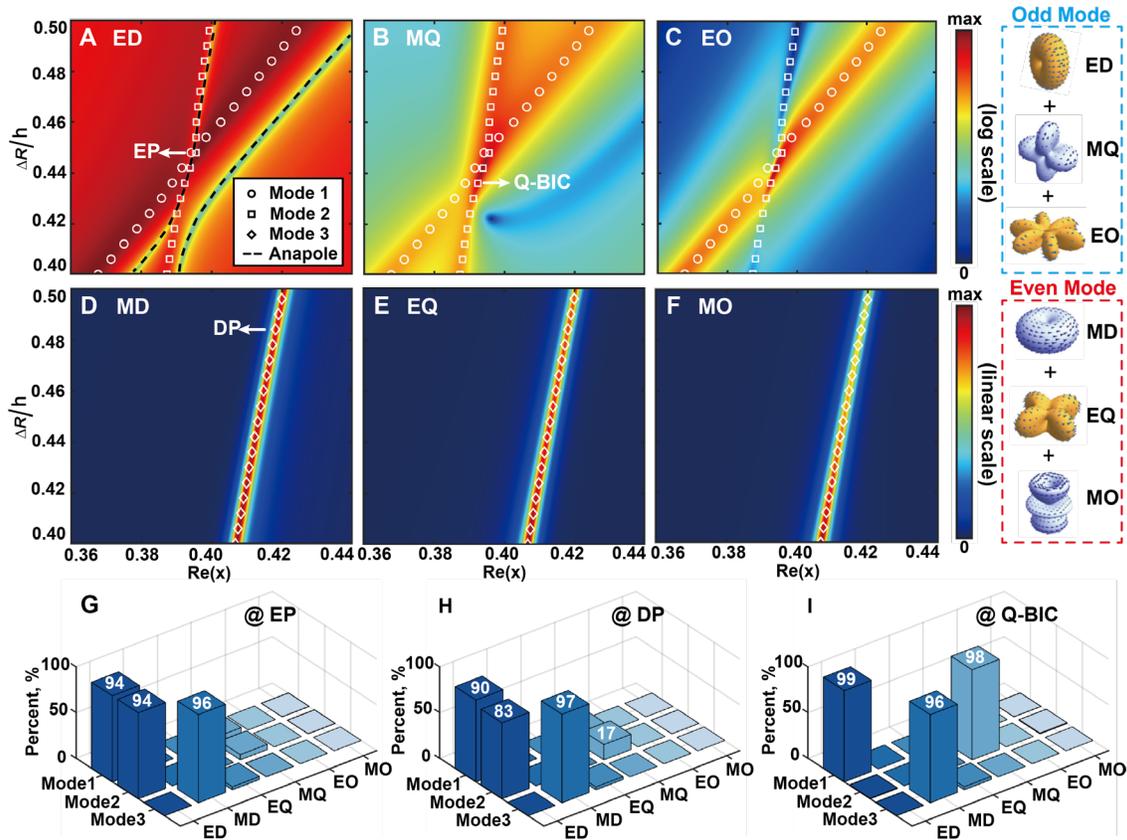

**Fig. 3. Multipole analysis.** The ED (**A**), MQ (**B**), EO (**C**), MD (**D**), EQ (**E**), MO (**F**) contribution of the scattering cross section excited by the normal incident wave with the dependency of aspect ratio $\Delta R/h$ and size parameter $x$. The various types of white dots indicate three simulated eigenmodes. The black dash lines indicate the position of scattering dark states in ED radiation channel. The right panels show that multipoles can be divided into odd (blue) and even (red) multipolar in an inversion symmetric particle. **(G-I)** The proportion of multipolar at different singularities: **(G)** EP, **(H)** DP and **(I)** Q-BIC.

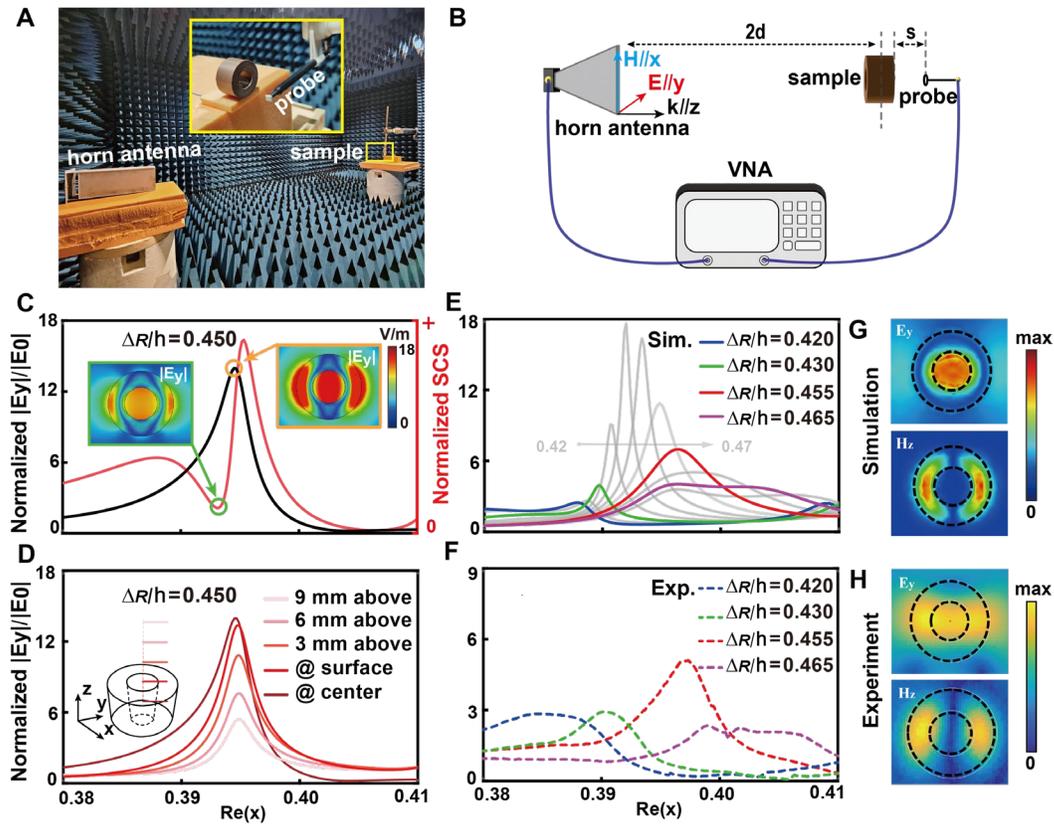

**Fig. 4. Near-field numerical and experimental results. (A)** Near-field experimental setup and **(B)** schematic illustration. **(C)** Calculated normalized scattering cross section and electric-field intensity enhancement in the center of ring-shaped resonator. The insets show the y-component of electric-field distribution at dip of the scattering cross section and peak of the electric-field enhancement. **(D)** Evolution of the y-component of the electric-field enhancement with the observation points changed in height. The simulated **(E)** and the corresponding experimental **(F)** results of the evolution of $|E_y|/|E_0|$ with continuous changing aspect ratio. The simulated **(G)** and experimental **(H)** profiles of y-component of electric field and z-component of magnetic field $3mm$ above the top surface of the resonator.

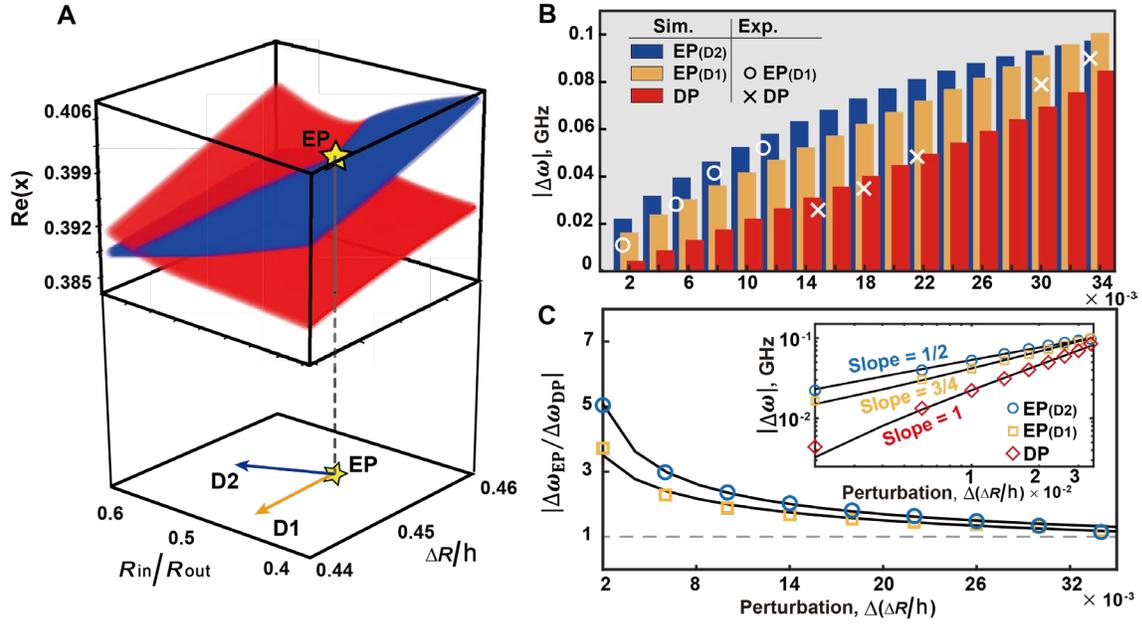

Fig. 5. Robustness of the system sensitivity based on eigenfrequency splitting. (A) The evolution of the real part of the eigenvalues, where golden star indicates the position of EP. The lower panel indicates two directions (D1: direction of PT symmetry breaking, D2: direction when $R_{in}/R_{out}$ is fixed) of variation of the structural parameters around EP. (B) Histograms showing the eigenfrequency splitting in ring-shaped resonator described by a comprehensive structural parameter at DP and EP sensors. The white circles and crosses represent the experimental data. (C) Dependence of the sensitivity enhancement $|\Delta\omega_{EP}/\Delta\omega_{DP}|$, the inset shows a double logarithmic plot between eigenfrequency splitting and perturbation at DP and EP, respectively.